\documentstyle[amsmath,amssymb,graphicx]{article}

\def\be{\begin{eqnarray}}
\def\ee{\end{eqnarray}}
\def\nn{\nonumber}

\def\l[{\phantom.[}

\def\MeV{\ \text{MeV}\ }

\hoffset= - 1.0in         

\begin{document}

\title{{\bf {Is pentaquark doublet a hadronic molecule?
}\vspace{.2cm}}
\author{{\bf A. Mironov$^{a,b,c,d}$}\ and \ {\bf A. Morozov$^{b,c,d}$}}
\date{ }
}

\maketitle

\vspace{-5.5cm}

\begin{center}
\hfill IITP/TH-07/15
\end{center}

\vspace{4.2cm}

\begin{center}

$^a$ {\small {\it Lebedev Physics Institute, Moscow 119991, Russia}}\\
$^b$ {\small {\it ITEP, Moscow 117218, Russia}}\\
$^c$ {\small {\it National Research Nuclear University MEPhI, Moscow 115409, Russia }}\\
$^d$ {\small {\it Institute for Information Transmission Problems, Moscow 127994, Russia}}\\
\end{center}

\vspace{1cm}

\begin{abstract}
A recently announced discovery by LHCb of a doublet of overlapping
pentaquark resonances poses a question of what can be the origin of this doublet
structure.
We attract attention to the fact that such degeneracy could naturally arise
if constituent "baryon" and "meson" were in the colored, rather than colorless states.
This is an appealing possibility, also because in such a case the pentaquark state would be
no less "elementary" than the other hadrons, and would provide a chance for
essentially new non-Abelian chemistry.
\end{abstract}

\vspace{1cm}

The search of exotic multiquark molecules is one of the main "side stories" in the history of QCD.
There is no clear reason why they can not exist, still they avoid experimental discovery for years.
Especially famous is pentaquark, which was predicted theoretically \cite{Bag,Diak} and almost found
experimentally \cite{pentaexpold}, and since then was many times closed and re-discovered \cite{Hicks},
leaving the issue in a very controversial form.

Recently a new seemingly reliable (at the level of $9\sigma$) observation of pentaquark
was reported by LHCb in CERN \cite{pentaexp}.
Namely, a decay mode was found $\Lambda^0_b=bud \longrightarrow  (s\bar d) + (c\bar c uud)$,
moreover, the pentaquark consists of two nearly overlapping resonances:
\be\label{pentamass}
J^P = 3/2^-: &   M_-= 4380\pm 37 \MeV, &
\Gamma_- = 205 \pm 104 \MeV \nn \\
J^P = 5/2^+: & M_+=4449.8\pm 4.2 \MeV, & \Gamma_+= 39\pm 24 \MeV
\ee
In fact, the assignments $(3/2^+,5/2^-)$ and $(5/2^+,3/2^-)$ are also experimentally acceptable though are less plausible.

From the most naive point of view these pentaquarks look like molecules made of
charmonium excitations and a proton, or of open charm meson and a corresponding baryon.
However, this is not the only option: the two (3-quark and quark-antiquark) constituents can actually
be in colored states, and then such configuration is not really a composite of two
anyhow-existing particles.

In this letter, we present some very naive arguments that experimental data can be pointing in this
{\it double}-exotic and therefore additionally exciting direction.

Note that arguments that could support an existence of pentaquarks have been considered for many decades within various dynamics frameworks, from the quark and bag models \cite{Bag} to the chiral Lagrangians \cite{Diak} (for a post-\cite{pentaexp} papers see \cite{DSigma,Ch}). Moreover, most of those
attempts dealt with a pentaquark made of light quarks (see, however, \cite{Lipkin}), and only recently pentaquarks with heavy quarks became more popular, see a review and references in \cite{Zou}, see also \cite{Oset}.
The charmed pentaquark made of heavy quarks is much more theoretically comfortable since thinking in terms of bound states is more justified in this case. Here, however, we would like to leave the dynamical side of the story and to concentrate more on color structures in order to understand to what extend one can rely on these two pentaquarks made of color singlet constituents.

In what follows, we briefly consider four obvious possibilities to make pentaquark of two constituents: the first two are from ordinary mesons and baryons (molecules), the second two from their colored analogues. The first two options are widely discussed in the literature, but we argue that they are not too nicely consistent with the newly discovered doublet structure. The second two options, though far more difficult to analyze, can have better chances.

\bigskip

\paragraph{1.} {\bf First} of all, the  masses of the candidate constituent mesons and baryons and mesons
from proton and charmonium families are:
\be
M_{uud} =  938\ MeV\ (p), \ \  1232\ MeV \ (\Delta^+), \ \ldots
\ee
and the charmonium $S$-states are
\be
\begin{array}{c|c|c|c|c}
&\eta_c(1S)&J/\psi(1S)&J/\psi(2S)&\eta_c(2S)\\
\hline
&&&&\\
M_{c\bar c} &  2980 + i\cdot 32\MeV&3097  + i\cdot 0.093\MeV&3686  + i\cdot 0.2 \MeV&3654 + i\cdot 11 \MeV
\end{array}
\ee
while the charmonium $P$-states are
\be
\begin{array}{c|c|c|c|c}
&\chi_{c0}(1P)&\chi_{c1}(1P)&\chi_{c2}(1P)&h_c(1P)\\
\hline
&&&&\\
M_{c\bar c} &  3415+i\cdot 10 \MeV &3511+i\cdot 0.8 \MeV &3556+ i\cdot 2 \MeV &3525 + i\cdot 0.7 \MeV
\end{array}
\ee

We remind that $\eta_c$ have the quantum numbers $J^{PC}$: $0^{-+}$ , $J/\psi$ -- $1^{++}$,
while $\chi_{c0,1,2}$ have $(0,1,2)^{++}$ and $h_c(1P)$ -- $1^{+-}$ (for mesons like $c\bar c$ the two parities $P$ and $C$
are equal to $(-)^{l+1}$ and $(-)^{l+s}$ respectively).

First of all, note that the spins of pentaquarks imply that 3/2 state can contain $\eta_c$ only in combination with $\Delta^+$ that has spin 3/2, while the spin 5/2 pentaquark is associated either with $(J/\psi,\chi,h_c)+\Delta^+$ or with $\chi_{c2}+p$.

Clearly, simple addition of masses does not provide anything close to required $M_\pm$.
In addition, from this pure combinatorial point of view,
the most natural option would be the lowest $P$-states of charmonium plus proton,
but then the sum of masses seems to exceeds $M_-$, which, however, has a bigger width than the state
with mass $M_+$.

Note, that if binding energy is significantly negative, as required by the mass differences, then the observed widths provide a
puzzle themselves:
the widths of all potential components are significantly smaller than
that of the pentaquarks, and there is no clear reason why this could be like this,
if we deal with a really bound state. In fact, this kind of possibility was studied in \cite{Oset}, and naturally led to much narrower states.

This is already enough to make the most naive charmonium+proton model not very plausible.

\paragraph{2.} {\bf Second}, for another constituent option one can consider open charm combinations. This option is more convenient for theoretical consideration, since the both constituents contain the heavy quark and, hence, the notion of bound state is better defined.

The open charm meson masses are
\be
M_{u\bar c} = 1865 \MeV\ (D_c^0); \ 2007 \MeV\ (D_c^{*0})\\
M_{d\bar c} = 1870 \MeV\ (D_c^-); \ 2010 \MeV\ (D_c^{*-})
\ee
\be
M_{cud} = 2286\MeV\ (\Lambda^+_c);\ 2453 \MeV\ (\Sigma_c^{+}); \ 2517.5 \MeV\ (\Sigma_c^{*+})\\ 
M_{cuu}=2454\MeV\ (\Sigma^{++}_c); \ 2518 \MeV\ (\Sigma_c^{*++})
\ee
where, as usual, the star means larger spin: $D$-mesons are scalars and $D^\ast$-mesons are vectors. Similarly, $\Lambda$'s and $\Sigma$'s have spin 1/2, while $\Sigma^*$'s have spin 3/2.

The combinations that could match the pentaquark quantum numbers are $D\Sigma^*$, $D^*\Lambda$, $D^*\Sigma$ and $D^*\Sigma^*$ for the spin 3/2 pentaquark and only the last of these for the spin 5/2 pentaquark. Now one needs to know what are the quantum numbers of the pentaquarks. For instance, the authors of \cite{DSigma} have chosen $(5/2^+,3/2^-)$ and presented the consideration of all coupled channels for $3/2^-$: all four from this option and $J/\psi p$ from the first option. They managed to reproduce the proper value of $M_+$ with a corresponding fine tuning, but left the story of the doublet aside.

Thus, this second option is not worse than the first one, but is not completed yet in describing the whole pentaquark doublet.

\paragraph{3.} The {\bf third} option is to look in another direction:
instead of composition of two colorless constituents one could compose two colored configurations.

Amusingly, then one naturally, if not unavoidably gets a doublet.
Indeed, in the very naive approximation
the wave function of the multiquark state
is a product of three: coordinate, spin and color.
A natural doublet appears in the color sector,
and again in the "baryon-meson" channel:
\be
\Box^3 \times (\Box\times\overline{\,\,\Box^{\phantom{5^5}\!\!\!}})
\supset 2\cdot [2,1] \times [2,1] \supset 2\cdot [0]
\label{colordeco}
\ee
(we use here the standard Young diagram notation for representations, in the usual QCD language
for $SU(3)$ group $\Box = [1] = 3$, $\overline{\,\,\Box^{\phantom{5^5}\!\!\!}} = \bar 3$, $[2,1]=8$).
The other singlet in this channel is just a product of singlets, which we discussed
(and rejected) as the first two options.

A serious weak point of the third option is that it "predicts" a too strong degeneracy:
the actual difference between the two states in (\ref{pentamass}) seems to be too big to fit into
this scenario. Indeed, the two $[2,1]=8$ in (\ref{colordeco}), which we suggested to associate as labeling
the two pentaquarks, have different symmetries only in the {\it two quark}
sector:  one $[2,1]$ comes from $[2]\times [1]$, another one from $[1,1]\times [1]$.
If the story was about the difference between two out of three quarks in
proton, then it does not seem  big enough to explain the different characteristics
of two pentaquarks.
Note, however, that
in the flavor sector one can still have all the three versions:
not only $(c\bar c)+(uud)$, but also $(u\bar c)+ (udc)$, $(d\bar c) + (uuc)$.
In the two latter cases the difference between components of the doublet
can be more pronounced than in the former one, hence, lifting.

\paragraph{4.} An alternative, {\bf forth} option can be then considered that the two components of the doublet are degenerate,
so that they were {\it not}
distinguished experimentally, and the two pentaquark states are internally colored doublet
and the singlet-singlet composite.
In variance with the third option, the difference between the two types of states is now significant: the color-charged constituents interact stronger than the color-neutral ones.
Still, this can be exactly what is needed to explain the difference between the two pentaquarks.
An even more pronounced "prediction" of this option is that
the multiplicities of the two pentaquarks can differ by a factor of two (since there are two
internally-colored and just one internally-neutral state).
If this was the case, it could strongly point towards {\it such} a possibility.

Note that in variance with option 2, here only {\it one} component of the pentaquark pair is made out of two colorless hadrons.
In other words, in this option the ordinary molecule should have mass $M_-$, in variance with \cite{DSigma}. This could explain why its width is higher: Van der Waals forces between colorless objects is clearly weaker than the Coulomb-like  between those in adjoint representation $[2,1]=8$.

Note also that some models with colored constituents were considered in the literature, but there was never a kind of a degenerate doublet, because the color of constituents was rigidly fixed.  For instance, considering the pentaquark as made from
two diquarks $qq$ and anti-quark \cite{dq}, one gets the only singlet, since the both diquarks are in representation [1,1]=$\bar 3$, while in order to get two more singlets one needs one of the quark pairs to be in the symmetric representation [2]=6.
Similarly, in the model of the diquark and triquark \cite{dtq} there is again only one color singlet, since the triquark
$qq\bar q$ is in the fundamental representation. In our options three and four the situation is different: we have {\it two different} triquark configurations in adjoint representation $[2,1]=8$ and, in option 4, also the third one in representation $[0]$.

\bigskip

Note that even the difference between the masses of two pentaquarks in (\ref{pentamass}) is not as small as it can seem.
At the first glance, different symmetries of coordinate function could strongly affect the masses, but instead the
mass dependence on spin can obey simple rules, if the two states lie on the same Regge trajectory. From this
point of view, such a "strong" dependence does actually occur: the typical slope of the Regge trajectory is about
${\Delta M^2\over \Delta J}\sim 1\ {\rm Gev}^2$,
and the reported difference $M_+^2-M_-^2 \approx 0.62 \pm 0.36\ {\rm Gev}^2 \sim 1\ {\rm Gev}^2$ is almost of the same magnitude.

\bigskip

To conclude, we provided some nearly evident analysis of the naive possibilities
to get a doublet of relatively narrow pentaquark resonances.
In variance with usual approaches to the problem, based on thorough study
of particular dynamical models,
we attempted to look at the qualitative explanations of the newly discovered qualitative effect:
existence of a somewhat peculiar doublet, nearly degenerate in mass, but quite different in spin
and decay width.
We claimed that the internal color structure of the would-be-constituents of the pentaquark,
which does not get enough attention
in the literature, can play a crucial role.
Not conclusively, but quite probably, this favors the option that the
pentaquarks are essentially new, intrinsically non-Abelian multiquark states,
{\it not} decomposable into a weakly coupled pair of hadrons.
In other words, the {\bf answer to the question in the title of this paper seems to be "NO"}.
Of course, much more work, both experimental and theoretical, is needed to really
justify or exclude this interesting possibility.

\section*{Acknowledgements}

This work was performed at the
Institute for Information Transmission Problems with the financial support of the Russian Science
Foundation (Grant No.14-50-00150).

\end{document}